\documentclass[showpacs,preprintnumbers,11pt]{revtex4}

\usepackage{graphicx}% Include figure files
\usepackage{dcolumn}% Align table columns on decimal point
\usepackage{bm}% bold math
\usepackage{amssymb,amsmath}

\begin{document}
%\preprint{hep-ph/0612xxx}
\title{High-energy Atmospheric Muon Flux Expected at India-Based \\ Neutrino Observatory}
%\title{Angular Distribution of Very High Energy Muon Flux at India-Based Neutrino Observatory}
\author{Sukanta Panda$^{a\ast}$ and Sergei I. Sinegovsky$^{b\dag}$}
 \vskip 1cm
\affiliation{{$^a$\em Departamento de Fisica Teorica C-XI and Instituto de Fisica Teorica IFT-UAM/CSIC C-XVI,}\\
{\em  Universidad Autonoma de Madrid, Cantoblanco, E-28049 Madrid,
Spain}\\
{$^b$\em Irkutsk State University, 664003 Irkutsk, Russia}\\
{{\em E-mail: }$^{\ast}$sukanta@delta.ft.uam.es,
$^{\dag}$sinegovsky@api.isu.ru}}
%\date{\today}

%\textheight 9in \textwidth 6in

\begin{abstract}
We calculate the zenith-angle dependence of conventional and prompt high-energy muon fluxes  at India-Based Neutrino Observatory (INO) depth. 
This study demonstrates a possibility to discriminate models of the charm hadroproduction including the low-x QCD behaviour of hadronic cross-sections relevant at very high energies.

\end{abstract}


%\keywords{Suggested keywords:} Cosmic ray interactions, Muons, Charmed 
%quarks \\

\maketitle

\section{Introduction}

 The cosmic ray (CR) spectrum is characterized by a sharply falling power law behaviour, $\frac{dN}{dE} \sim E^{-(\gamma + 1)} $\cite{gai}. The spectrum gets more steeper around
$10^6$ GeV with the spectral index $\gamma$ changing from $1.7$ to
$2.1$ - this region is called the {\it knee}. Around $E\sim 5\times
10^9$ GeV, one observes a flattening of the spectrum, with the
spectral index $\gamma$ falling between $1.4$ and $1.7$. 
This is the so called {\it ankle}. 
%The change in the slope of the spectrum at these two places is a puzzling issue.  
These two breaks of the primary spectrum are still open questions of CR physics.
The region beyond the ankle is the regime of ultra high energy cosmic rays. There is not much data available in that region and no clear consensus exists on the composition or the particle content in this region \cite{nag}. It is generally believed that the change in the slope around the knee is astrophysical in nature rather than any specific change in hadronic
properties and/or interactions~\cite{nag,kas3}. 
An overview of hadronic interactions and cosmic rays can be found in~\cite{lowxcr}.

The atmospheric muon flux originated from decays of pions and kaons is commonly called the {\it conventional muon} flux. There is expected rather sharp reduction of the 
conventional muon flux above a few TeV \cite{lip} due to the increasing decay lengths and decreasing interaction lengths of pions and kaons. Therefore, at very high 
energies the bulk of muons is expected to arise from the semileptonic decay modes of heavy shortlived hadrons, predominantly the charmed ones. This component is called the {\it prompt muons}. 
It is known that the prompt muon flux is only about $10\%$ smaller than the prompt
$\nu_{\mu}$ flux at the surface of the earth. Therefore, measurement of the atmospheric prompt muon (APM) flux at high energies will ensure a normalization for the atmospheric prompt neutrino (APN) flux, and a direct comparison of the two is both desirable and necessary.
This study is necessary because the atmospheric neutrino flux is unavoidable background to VHE neutrino experiments. 

There are sizeable uncertainties in theoretical predictions 
for the prompt lepton fluxes (see~\cite{bug89, costa01} 
for review). The reason is mainly due to the vastly 
different choices for the charm production cross-section -- perturbative QCD (pQCD) with 
a $K$ factor \cite{tig}, next-to-leading order (NLO) pQCD \cite{prs, ggv2}, phenomenological 
nonperturbative approach, such as the recombination quark-parton model (RQPM) or quark-gluon string model (QGSM)~\cite{bug89}. The experimental situation is not 
very precise either at this stage. Various experiments~\cite{lvd} 
provide upper limits on the APM fluxes in the energy range of interest, which allow a large variation in the prompt fluxes. One can therefore expect that better measurements of 
high-energy muon fluxes can play a definitive role in selecting the charm production models, and thereby, also providing invaluable information about parton densities at such low-$x$ and high energy values. Another related source of large theoretical uncertainties 
is strong dependence of the hadronic cross-sections on the renormalization and factorization scales. This is partly related to the naive extrapolation of parton distribution functions to very different energy and $x$-values. For the case of conventional fluxes 
originating from the pions and kaons, these issues are in much better control and 
therefore the predictions stand on a sound footing.

Earlier authors of Ref.~\cite{gp, mp} explored the possibility of utilizing the
high energy prompt muon flux(es) in order to investigate whether the
general expectations expressed above can in practice help in
selecting the charm production model/parameterization and also the importance of
the heavy composition of cosmic rays above knee. They chose some of the 
models often used and compare the predictions, incorporating the saturation model 
of Golec-Biernat and Wuthsoff~\cite{gbw}. However while esimating their
event rates of muons in a 50 kT Iron detector like INO one~\cite{ino} they did not consider 
the angular dependence of the muon fluxes at rock depth. 
Angular dependence of muon flux due to surrounding rock is really important 
for correct estimation of the muon event rate inside such a detector. 
%In this letter we show the angular dependence of the conventional muon flux as well as prompt one at INO rock depth.
In this work we calculate the high-enery AM flux, conventional as well as prompt, at INO rock depth taking into account the distortion in the surface muon zenith-angle distribution due to specific topography of the INO site. 

It is therefore quite clear from all these models that the lepton fluxes at the end are strongly sensitive to the charm production cross section. Till the knee, the cosmic ray flux and composition is rather established and therefore, the only source of large error is the charm cross section. This therefore gives us a unique possibility to gain information about heavy quark production mechanism at high energies and low $x$. 

%%%%%%%%%%%%%%%%%%%%%%%%%%%%%%%%%%%%%%%%%%%%%%%%%%%%%%%%%%%%%%%%

\section{Surface atmospheric muon flux and the calculation technique}
\subsection{Topography of PUSHEP site}

 The slant depth $X$ depends on the topography of the rock 
surrounding the INO detector. 
PUSHEP is the selected site for this purpose. One can assume a constant depth
which is equal to the vertical depth just above the cavern. The vertical
depth of PUSHEP site is 1.3 km of rock. Another assumption is that of a triangle topogarphy.
In this case the slant depth for given zenith angle $\theta$  is calculated as
\begin{equation}
X(\theta) = \frac{h_0}{\cos\theta + (h_0/l_0)\,\sin\theta} \  ,
\end{equation}
where $h_0=1.3$ km is the vertical depth, $l_0=2.1$ km is the half-length of the approach
tunnel and $\tan\omega=h_0/l_0$ is the slope of the mountain.
%which equals to 2.1 km. 
% In this case the slant depth is given by equation
\begin{figure}[!hb]
\begin{center}\hskip -0mm 
\includegraphics[width=7.0cm]{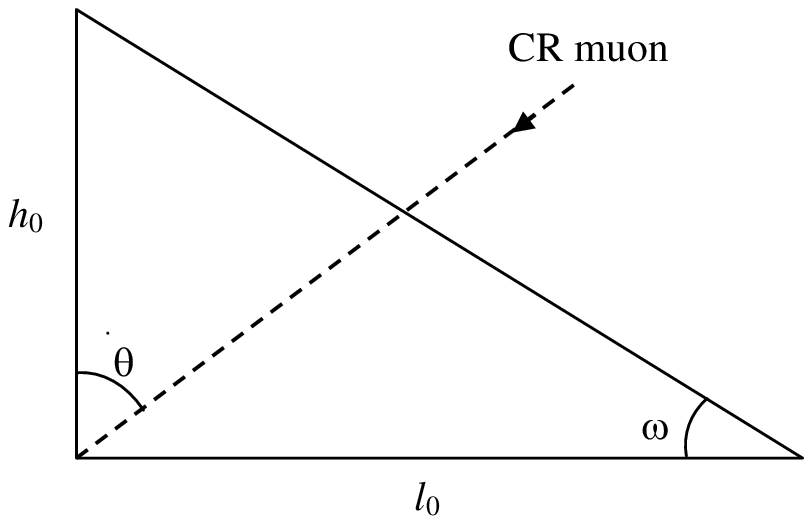} \hskip 5mm 
%\caption{Geometry of INO site.} 
\includegraphics[width=8.0cm]{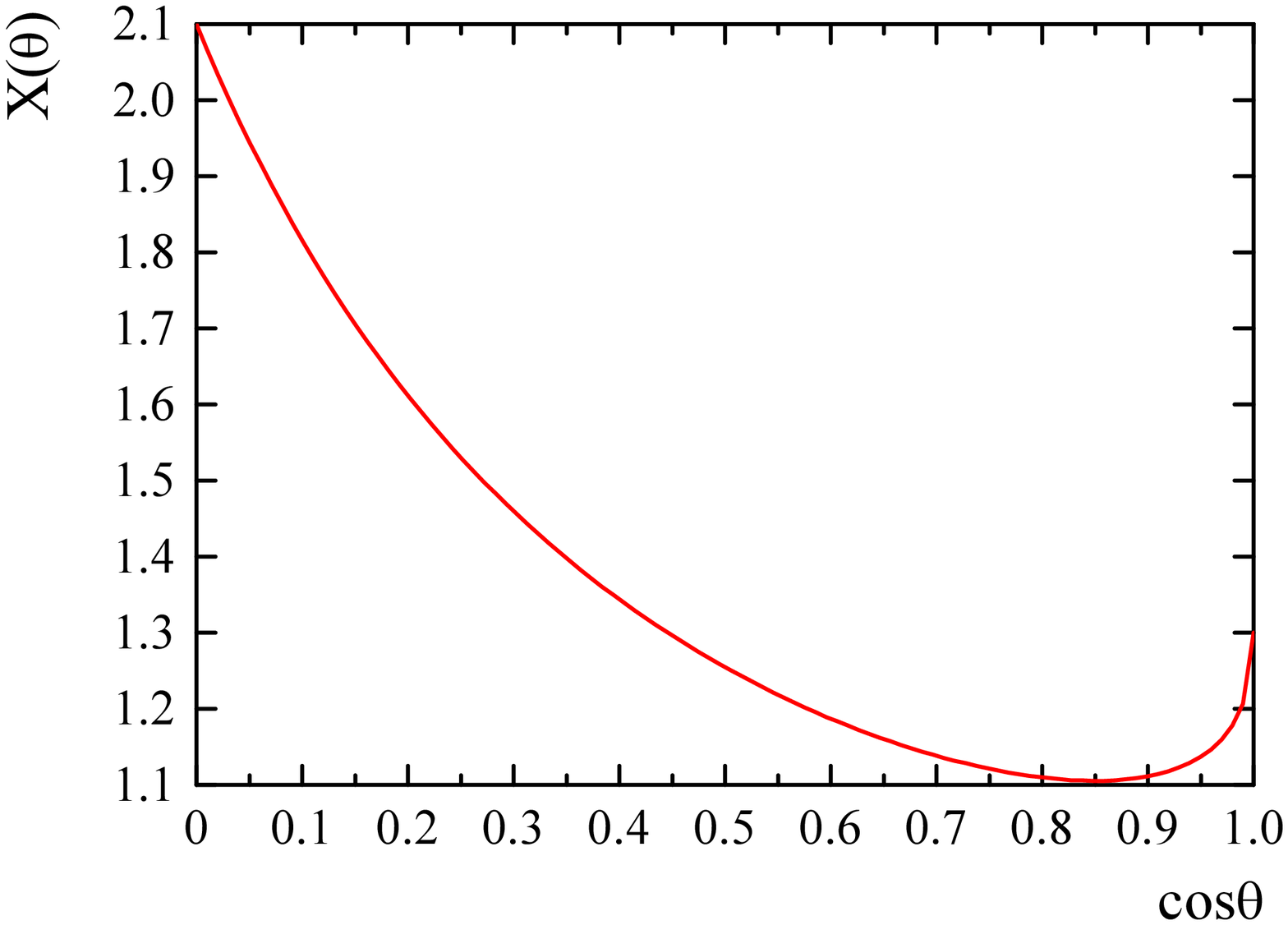} 
\end{center}
\hskip   0mm 
\vskip -10mm
\caption{Geometry and slant depth of PUSHEP site.}
\label{fig1}
\end{figure}
The triangle nature of site and the slant depth $X(\theta)$ are shown in Fig.~\ref{fig1}. 
For the rock density $\varrho$ we adopt here value $2.72$ g/cm$^{3}$.
The column depth $h(\theta)$  related to the slant depth, $h(\theta)= \varrho X(\theta)$, 
%\begin{equation}
%h(\theta)= \varrho X(\theta),
%\end{equation}
varies between $h_{\rm min}\simeq 3.0\cdot 10^5$ g$\cdot$cm$^{-2}$ ($3$ km w. e.) that %3.007
corresponds to $\cos\theta_{\rm m}\simeq 0.85$ and  % 5.712
$h_{\rm max}\simeq 5.71\cdot 10^5$ g$\cdot$cm$^{-2}$  ($5.71$ km w. e.) near horizontal. Near  vertical direction column depth is about 3.54 km w. e. 
%\begin{figure}[]
%\hbox{\hspace{0cm}
%\hbox{\includegraphics[scale=1.1]{geometry.eps}}}
%\caption{Geometry of INO site.}
%\label{fig1}
%\end{figure}
 
%%%%%%%%%%%%%%%%%%%%%%%%%%%%%%%%%%%%%%%%%%%%%%%%%%%%%%%%%%%%%%%%%%%

\subsection{Parameterization of the conventional muon spectrum at sea level}

%Muons and neutrinos are the most penetrating particles produced in the atmosphere
%by the cosmic ray interactions. The cosmic ray muon flux is known to
%fall same as their CR primary counterpart with a power law of $E^{-2.7}$.

The surface muon flux is rather well measured up to TeV and can be described by different analytical formulae taking into account the zenith-angle dependence. Here we list some of them which were used in present calculations.
First of all we use Gaisser's muon flux parameterization~\cite{gai,gaisser04} (in inits of 
$ \rm{cm}^{-2}\rm s^{-1}\rm{sr}^{-1}\rm{GeV}^{-1}$)
\begin{equation}\label{gais} 
\phi^{\rm \pi, K}_{\mu}(E_{\mu}) =0.14E_{\mu}^{-2.7}\left[\frac{1}{1+1.1(E_{\mu}/115 {\rm GeV})\cos\theta}+ \frac{0.054}{1+1.1 (E_{\mu}/850 {\rm GeV})\cos\theta}  \right]. 
%  , \  \rm{cm}^{-2}\rm s^{-1}\rm{sr}^{-1}\rm{GeV}^{-1}.
\end{equation} 
For our purpose  we work with a modified muon flux formula obtained by Tang et al.~\cite{tang}.

%As other cases we consider Bugaev et al.~\cite{bug98} and  Reyna's
Next parameterization of the conventinal muon flux  we use here is that  by Bugaev et al.~\cite{bug98} for vertical direction:
\begin{equation}\label{fit1}
\phi^{\rm \pi, K}_{\mu}(p_{\mu}, 0^\circ) =
Cp_{\mu}^{-(\gamma_0+\gamma_1 z+\gamma_2 z^2+\gamma_3 z^3)} \, , \   \rm{cm}^{-2}\rm s^{-1}\rm{sr}^{-1}\rm{(GeV/c)}^{-1},
\end{equation}
where $z=\log_{10}(p_{\mu}/1~\rm{GeV/c})$. Values of parameters 
in~Eq.~(\ref{fit1}) are listed in Table~\ref{t2} for different 
momentum ranges. 
The muon energy spectrum is $\phi^{\rm \pi, K}_{\mu}(E,\theta) = 
(E_{\mu}/p_{\mu})\,\phi^{\rm \pi, K}_{\mu}(p_{\mu}, \theta) $. 
\begin{table}[!h] %%%%%%%%%%%%%%%%%%%%%%%%%%%%%%%%%%%%%%%%%%%%%%%%%%%%%%
\protect\caption{Parameters in Eq.~(\protect\ref{fit1}) for the vertical
                 energy spectrum of conventional muons at sea level.
\label{t2}}
\center{\begin{tabular}{lccccc}\hline\hline
Momentum range, GeV/c $\ $ &
                 $C$, (cm\,${}^{2}$s\, sr\, GeV/c)$^{-1}$
       & $ \gamma_0  $ & $ \gamma_1$ & $\gamma_2$ & $\gamma_3$ \\\hline
$1.0 - 927.65$  & $2.950\times10^{-3}$
       &   0.3061   &   1.2743   &   -0.2630  &   0.0252  \\
$927.65 - 1587.8$  & $1.781\times10^{-2}$
       &   1.7910   &   0.3040   &    0       &   0        \\
$1587.8 - 4.1625\times10^5$  & $14.35$
       &   3.6720   &   0        &    0       &   0        \\
$ > 4.1625\times10^5$                     &            $10^{ 3}$
       &   4.0        &   0        &    0       &   0 \\
\hline\hline
\end{tabular}}
\end{table} %%%%%%%%%%%%%%%%%%%%%%%%%%%%%%%%%%%%%%%%%%%%%%%%%%%%%%%%
For inclined directions we use  zenith-angle dependence given in 
Ref.~\cite{tanya} (see also~\cite{ts}). As the third 
parameterization of the atmospheric muon flux  we use the formula 
given in Ref.~\cite{reyna}.

%%%%%%%%%%%%%%%%%%%%%%%%%%%%%%%%%%%%%%%%%%%%%%%%%%%%%%%%%%%%%%%%%%%

\subsection{Prompt muon contribution} %fraction

\begin{figure}[!t]
\begin{center}
\includegraphics[width=9.0cm]{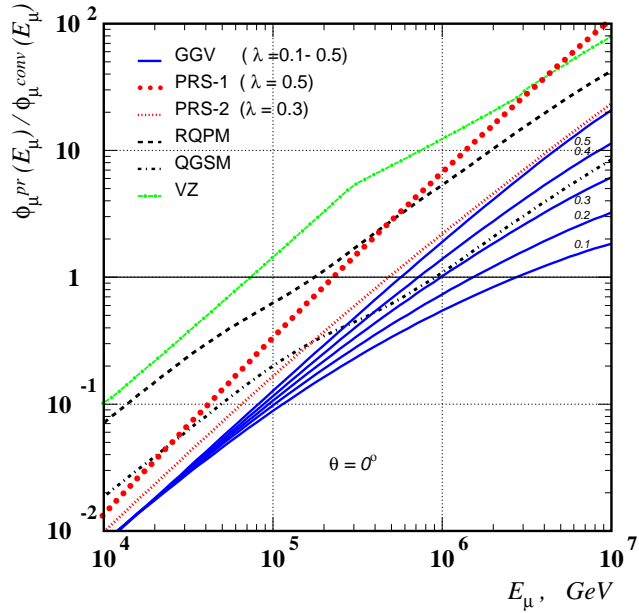} 
\end{center}
\vskip -15mm
\hskip   0mm 
\caption{Ratio of the prompt muon flux to the conventional one at ground level.}
\label{cc-rat}
\end{figure}

Atmospheric prompt muon flux predictions are reviewed in Refs.~\cite{bug89, costa01}.
Ratios of the differential energy spectra of muons at sea level originated 
from charmed particle decays to that of ($\pi,\,K$)-decays (conventional muons) calculated 
for a variety of charm production models are shown in Fig.~\ref{cc-rat}  (see also~\cite{misaki}). Here  PRS stands for the model~\cite{prs}, GGV for ~\cite{ggv2}, RQPM and QGSM for~\cite{bug89, bug98}, and  VZ for Volkova and Zatsepin~\cite{vz}).
Among them we dwell below on quark-gluon string model (QGSM), 
as a sample of phenomenological nonperturbative approach, and also on some of models based on perturbative QCD computations,  GGV~\cite{ggv2} and GBW~\cite{gbw}. 

Gelmini, Gondolo and Varieschi (GGV)~\cite{ggv2} have included NLO corrections for the charm production with $ xg(x) \sim x^{-\lambda},$  ($\lambda$ varying in the range $0-0.5$). These
results obey the following parameterization for the sea-level muon fluxes                    (see also~\cite{misaki}):
\begin{equation}\label{ggv}
\phi^{\rm GGV}_{\mu}(E_{\mu}) =
A\left(\frac{E_{\mu}}{1~\rm{GeV}}\right)^{-(a+by+cy^2+dy^3)} , \  
 \rm{cm}^{-2}\rm s^{-1}\rm{sr}^{-1}\rm{GeV}^{-1}.
\end{equation}
where $y= \log_{10}(E_{\mu}/1~\rm{GeV})$. The parameters are given in the
Table~\ref{t1}. We choose two representative sets corresponding to
$\lambda=0.1$ (GGV01) and $\lambda=0.5$ (GGV05).
\begin{table}[!hb]
\protect\caption{GGV  parameters  for the prompt muon fluxes.
% MRST PDFs are adopted and the scale is set to $\tilde{M}=2\tilde{\mu}=2 E_T = 2 (k_T^2 + m_c^2)$;  $m_c=1.25$ GeV and  $k_T$ is the transverse momentum of charm quark
\label{t1}}
\center{\begin{tabular}{l|ccccccr|} \hline\hline
Model  & A, cm$^{-2}$s$^{-1}$sr$^{-1}$GeV$^{-1} \ $&  a   &  b  &  c  &  d \\\hline 
GGV01& $3.12\times10^{-6}$\,& 2.70 & -0.095 & $1.49\times 10^{-2}$& $-0.2148\times 10^{-3}$ \\
GGV05& $0.58\times10^{-6}$ & 1.84 & 0.257 & $-4.05\times 10^{-2}$ & $2.455\times 10^{-3}$ \\
\hline\hline 
\end{tabular}}
\end{table}

QGSM flux parameterization (that is valid for $\theta \lesssim 80^\circ$) may be 
written~\cite{bug89} as
\begin{equation}\label{qgsm}
\phi^{\rm QGSM}_{\mu}(E_{\mu}) =1.09\cdot 10^{-18}\left(\frac{E_{\mu}}{100\,\rm{TeV}}\right)^{-3.02}
\left[ 1+  \left(\frac{E_{\mu}}{100\,\rm{TeV}}\right)^{-2.02} \right]^{-0.165}, \  
 \rm{cm}^{-2}\rm s^{-1}\rm{sr}^{-1}\rm{GeV}^{-1}.
\end{equation}

As last representative model, we consider flux calculation
within the saturation model proposed by Golec-Biernat and Wuthsoff  \cite{gbw}. For this model, we consider two cases~\cite{mp}): GBW1, where the protons are taken to be the primary, and GBW2, where we include the effect of heavy elements also. 
The sea level prompt muon flux due to  GBW1 and  GBW2 can be parameterized as Eq.~(\ref{GBW1_pm}) and Eq.~(\ref{GBW2_pm}) respectively:
\begin{equation}\label{GBW1_pm}
\phi^{\rm GBW1}_{\mu}(E_{\mu}) =2.35\cdot 10^{-8}
\left(\frac{E_{\mu}}{1~\rm{GeV}}\right)^{-2.17145-0.04984 y},  \  
 \rm{cm}^{-2}\rm s^{-1}\rm{sr}^{-1}\rm{GeV}^{-1},
\end{equation}
\begin{equation}\label{GBW2_pm}
\phi^{\rm GBW2}_{\mu}(E_{\mu}) =1.09\cdot 10^{-8}
\left(\frac{E_{\mu}}{1~\rm{GeV}}\right)^{-1.79371-0.10873 y} , \  
 \rm{cm}^{-2}\rm s^{-1}\rm{sr}^{-1}\rm{GeV}^{-1}.
\end{equation}
These two cases are different in nature, with the expectation that GBW2 should lead to a decreased muon flux at higher energies.

%%%%%%%%%%%%%%%%%%%%%%%%%%%%%%%%%%%%%%%%%%%%%%%%%%%%%%%%%%%%%%%%%%

\subsection{Method to calculate the muon flux under thick layer of the rock}

In these computations we base on the semianalytical method for the solution of muon transport equation stated in  Ref.~\cite{nsb94} (see also ~\cite{bug98, ts}).
The method allows to consider real atmospheric muon spectrum and the energy behavior of discrete energy loss spectra due to radiative and photonuclear interactions of muons in   matter. Only ionization energy loss of muons are treated as continuous one. The method provides effective tool to compute the energy spectra of cosmic-ray muons at large depths of homogeneous media. The benefits of this approach are to carry out verifications of the primary CR spectrum and composition, charm production models, models of the photonuclear interaction with high performance and good precision. This  enables to estimate the sea-level muon spectrum using the data of underground/underwater measurements evading the difficult inverse scattering problem.

%%%%%%%%%%%%%%%%%%%%%%%%%%%%%%%%%%%%%%%%%%%%%%%%%%%%%%%%%%%%%%%%%%

\section{Expected muon flux at the depth of PUSHEP site} %of the INO detector
%\section{Results}

\begin{figure}[ht!]
\begin{center}
\includegraphics[width=8cm]{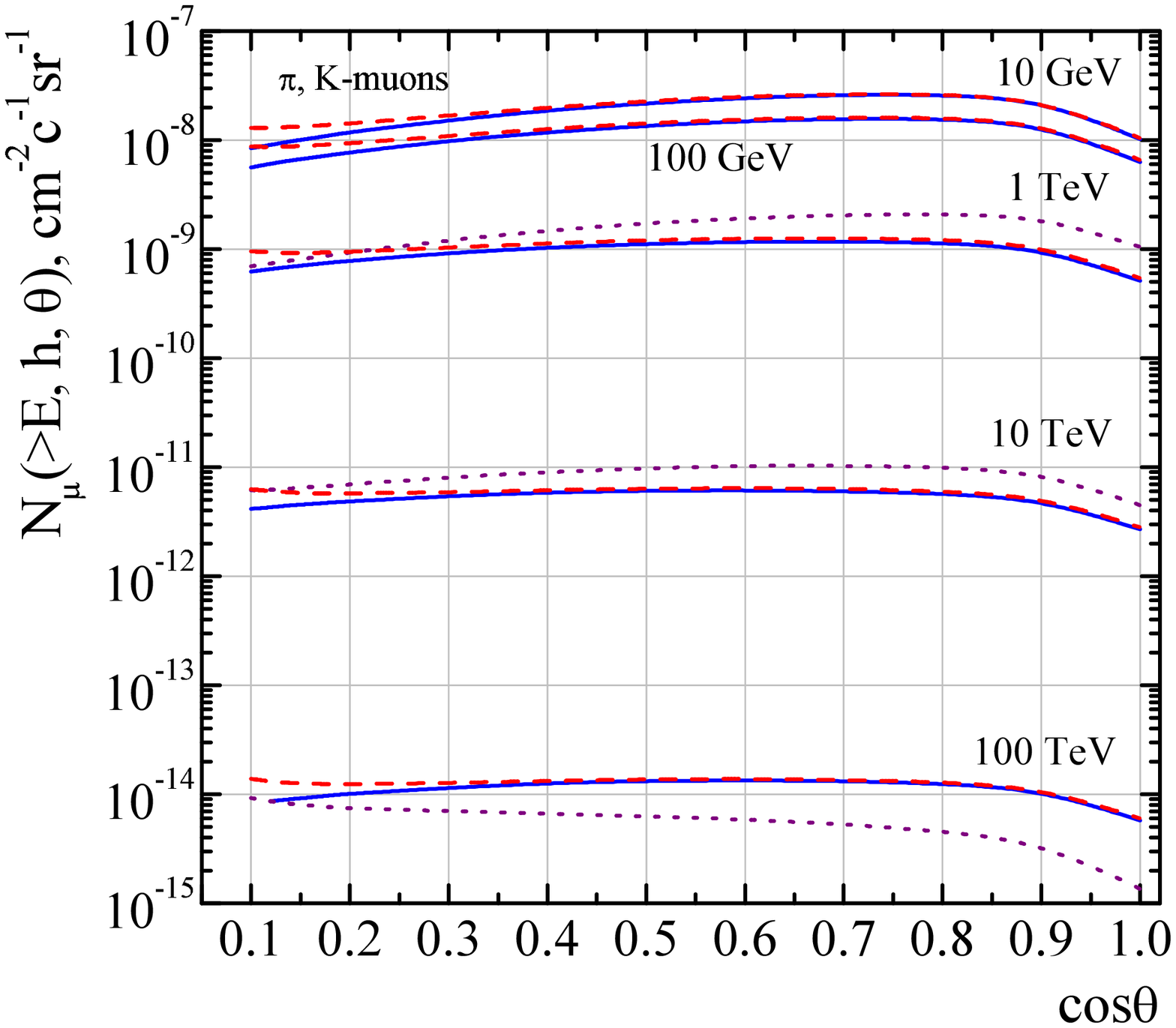}
\hskip   0mm 
\end{center}
\vskip -5mm
\caption{Angle distributions of the conventional muons near the INO detector.}
\label{conv_ang}
\end{figure}

Zenith-angle distributions of the conventional muon flux calculated for five values of the minimal energy of muons in the range $10$--$10^5$ GeV  at  depth $1.3$ km of INO detector are shown  in Fig.~\ref{conv_ang}. Here solid lines represent computations for the surface muon spectrum~\cite{bug98} by Bugaev et al. with usage of the angle dependence obtained in Ref.~\cite{tanya} (see also~\cite{ts}).
Dashed lines, almost superimposed on solid ones but near horizontal 
directions, show results for the spectrum by Tang et al.~\cite{tang} whereas dotted ones show 
that for the spectrum by Reyna~\cite {reyna}. The geometry of the INO site is 
reflected in the flat shape of the underock distribution(see Fig.~\ref{fig1}). 
%--------------------------------------------------------
\begin{figure}[ht!]
\begin{center} %\vskip 20mm
\includegraphics[width=8cm]{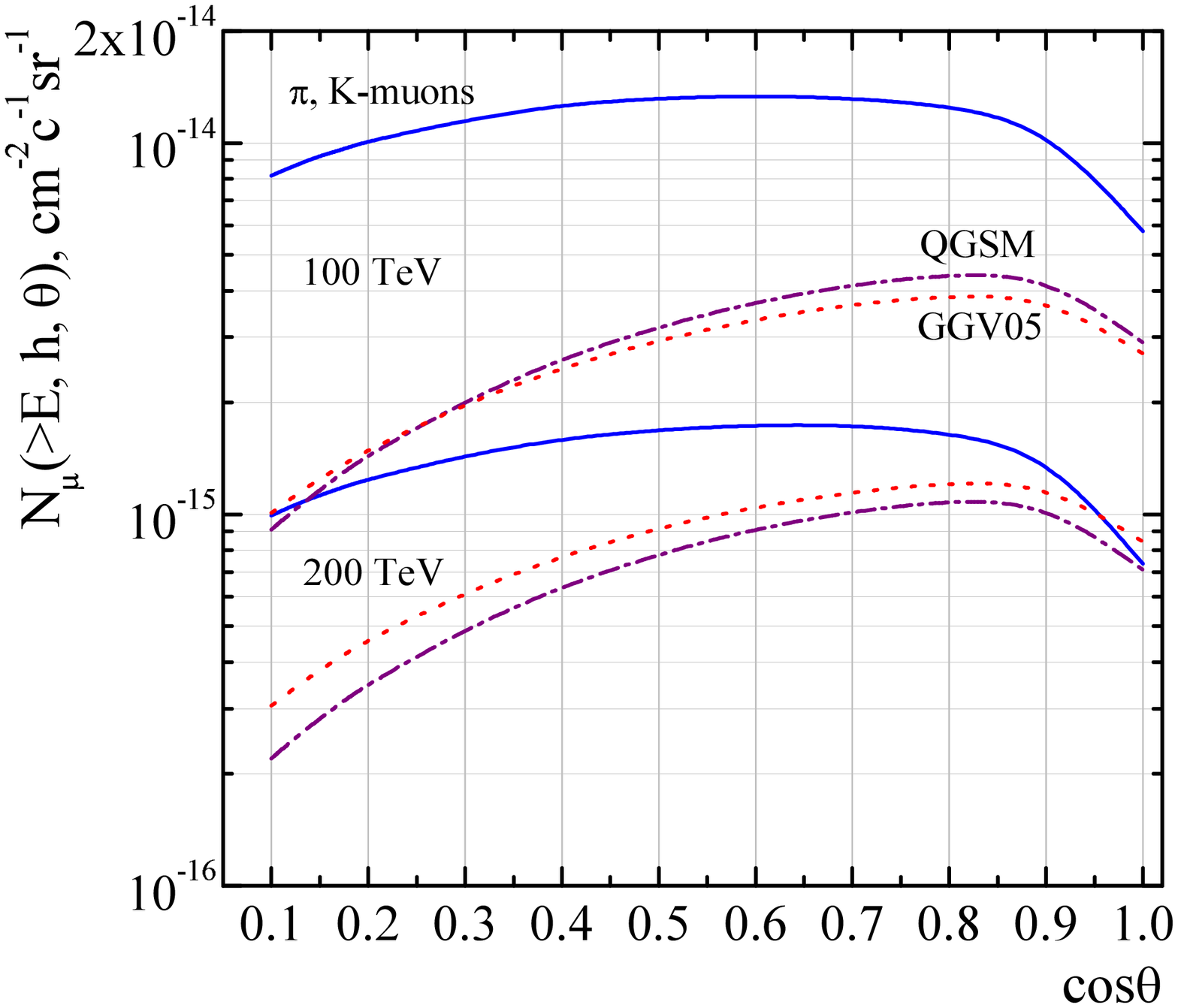}
\hskip   0mm 
\end{center}
\vskip -5mm
\caption{Zenith-angle distributions of atmospheric muons at $100$ and $200$ TeV.}
\label{100TeV}
\end{figure}       
Zenith-angle dependence of the conventional and prompt muon fluxes at the INO depth are shown in Figs.~\ref{100TeV} and \ref{500TeV}. In Fig.~\ref{100TeV} are shown the prompt muon flux at muon energy above  $100$ and  $200$ TeV calculated  with QGSM charm production cross sections (dash-dotted lines) and that of GGV models.  For muon energy above $500$ TeV we also plot predictions obtained for GBW model (dashes line in Fig.~\ref{500TeV}).  
As one may clearly observe in Fig.~\ref{500TeV}, measurement 
of high muon flux near the vertical at INO depth 
could allow to discriminate between GGV01($\lambda=0.1$) model and GGV05($\lambda=0.5$) or  QGSM one. While the GBW prompt muon flux is unlikely to be observed at 500 TeV. 

Differential muon spectra (left panel) at INO depth and integral ones (right panel)  are 
presented in Fig.~\ref{spectra} for $\cos\theta=0.7$, where solid line shows the conventional muon flux obtained  with usage of Bugaev et al. boundary spectrum and circles denote that for Gaisser's spectrum.  We can see in Fig.~\ref{spectra} that crossover  energy for the conventional muon flux  and the GGV05 prompt one  is about $300$ TeV, therefore   
it seems that more suitable for the prompt muon identification is to analyse the zenith-angle dependence of high-energy muon flux (see Fig.~\ref{100TeV}).
%%%%%%%%%%%%%%%%%%%%%%%%%%%%%%%%%%%%%%%%%%%%%%%%%%%%%%%%%%%%%%%%%%%%%%%%%%%%%%%%
%--------------------------------------------------------
\begin{figure}[ht!]
\begin{center} %\vskip 20mm
\hskip   0mm 
\includegraphics[width=8cm]{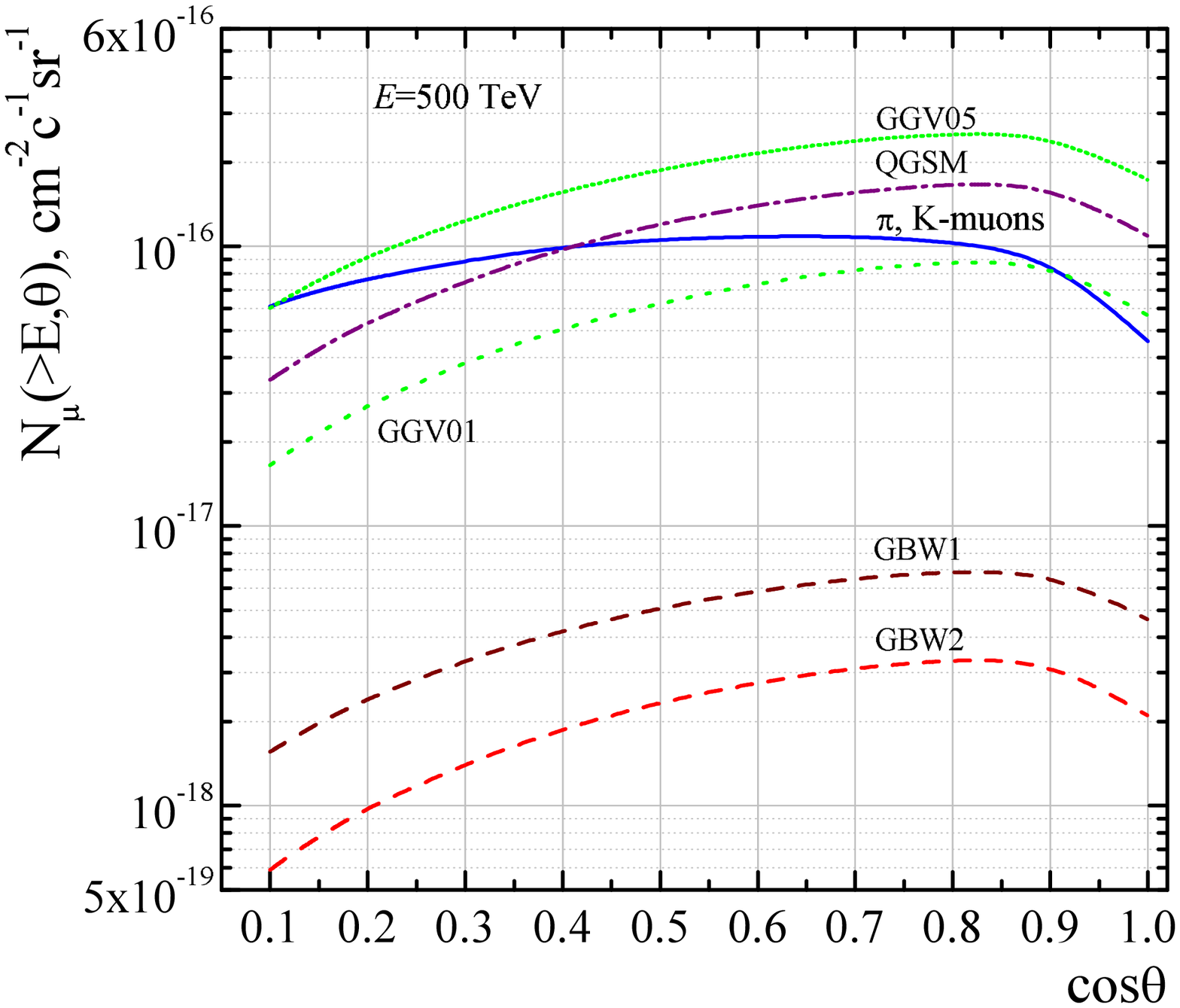}
\end{center}
\vskip -5mm
\caption{Very high-energy zenith-angle distributions of atmospheric muons.}
\label{500TeV}      
\end{figure}
%--------------------------------------------------------
\begin{figure}[ht!]
\begin{center} \hskip   -5mm
\includegraphics[width=7cm]{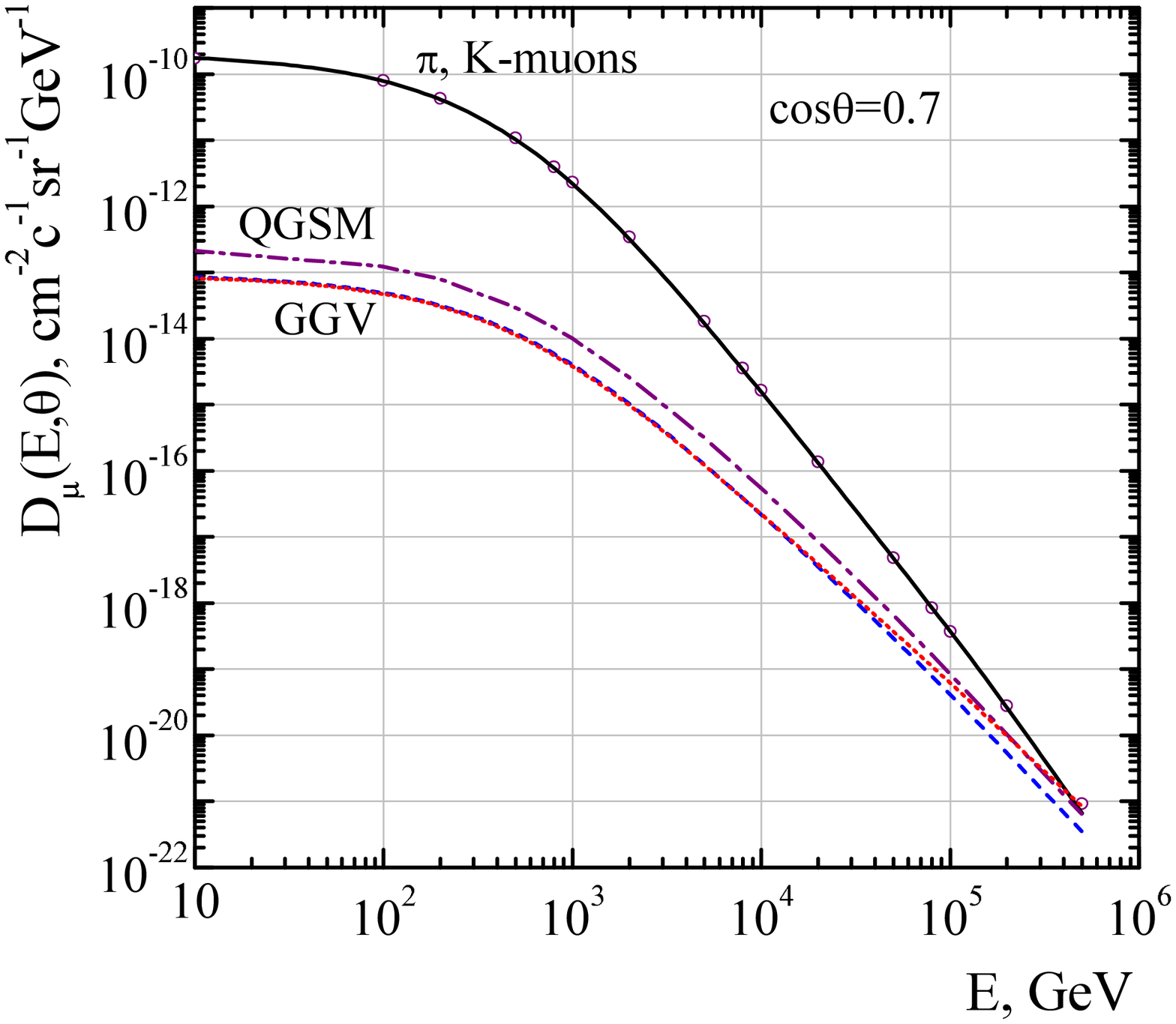}
\hskip   10mm 
\includegraphics[width=7cm]{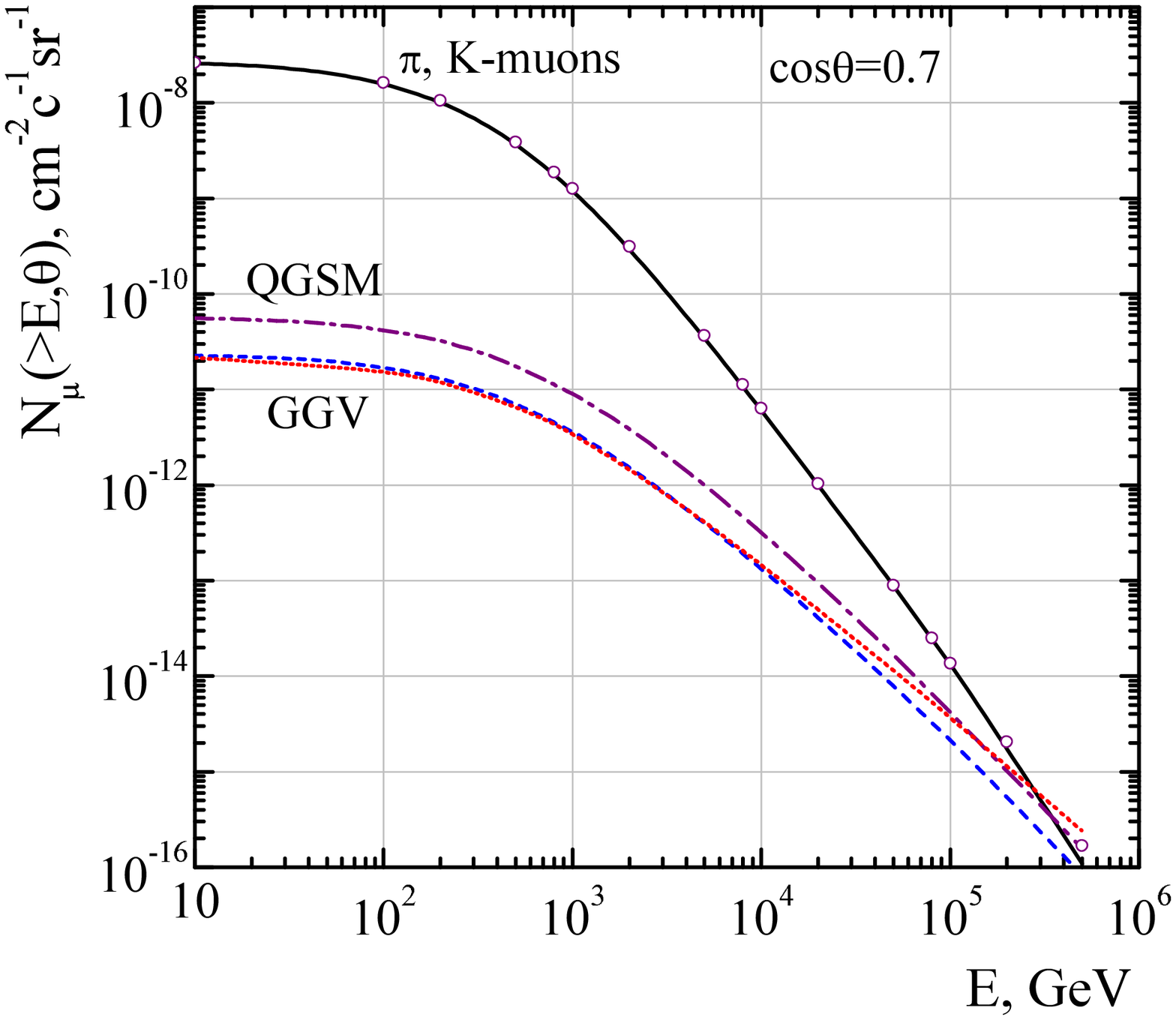}
\end{center}
\vskip -5mm
\caption{Energy spectra of the atmospheric muons near the INO detector.}
\label{spectra}
\end{figure}
%--------------------------------------------------------

Number of the muon events per steradian per year expected at INO detector  near direction $\cos\theta =0.7$ is presented in Table~\ref{t3} (see details in Ref.~\cite{gp}). 
\begin{table}[!h] %%%%%%%%%%%%%%%%%%%%%%%%%%%%%%%%%%%%%%%%%%%%%%%%%%%%%%
\protect\caption{Number of the muon events per steradian per year expected at INO detector. \\  %       near direction $\cos\theta =0.7$.
\label{t3}}
\center{\begin{tabular}{c|cccc|ccc}\hline\hline
$E_\mu$, TeV   & $\quad $ conv.  $\quad $   &  GGV01 $\quad $ & GGV05 $\quad $ & QGSM $\quad $  & $R_c^{\rm GGV01}$ & $R_c^{\rm GGV05}$ & $R_c^{\rm QGSM}$\\\hline
$10  $  & $60097 $ & $1235 $   & $ 1353 $   & $3037 $  & $0.02$ & $0.022$ & $0.05$  \\
$50  $  & $832 $ & $73 $   & $ 105 $   & $159 $   & $0.087$ & $0.126$ & $0.19$ \\
$100 $  & $132 $ & $20 $   & $ 34 $   & $41  $  &$0.15$  & $0.258$ & $0.31$   \\
$200 $  & $ 20$ & $5.0 $   & $ 10 $   & $10  $  &$0.25$  &  $0.50$ & $0.50$  \\
$300 $  & $ 6.0$ & $2.0 $   & $ 5.0 $   & $4.0  $ & $0.33$  &  $0.83$ & $0.66$    \\
$400 $  & $ 2.6$ & $1.0 $   & $ 2.6 $   & $2.2  $   & $0.38$  & $1.0$ & $0.85$  \\
$500 $  & $ 1.4$ & $0.6 $   & $ 1.6 $   & $1.2  $  & $0.43$ & $1.14$  & $0.86$  \\
\hline\hline
\end{tabular}}
\end{table} %%%%%%%%%%%%%%%%%%%%%%%%%%%%%%%%%%%%%%%%%%%%%%%%%%%%%%%%
%%%%%%%%%%%%%%%%%%%%%%%%%%%%%%%%%%%%%%%%%%%%%%%%%%%%%%%
Last three columns in Table~\ref{t3} represent the ratio of the conventional muon flux to 
the prompt muon one due to three charm production models, GGV01, GGV05  and  QGSM, respectively.

\section{Summary}
%\section{Discussion}
The shape of zenith-angle distributions of conventional muons is nearly flat 
(see Figs.~\ref{conv_ang}--\ref{500TeV}). Therefore muons arriving at the detector close to vertical directions are  more favorable to measure the prompt muon flux.
The prompt muon contribution to the atmospheric muon flux 
increases with energy because of lower value of the energy spectrum index. 
The  ``crossover'' energy, $E_c$, at which the prompt muon flux 
becomes equal to the conventional one, depends strongly on 
the charm production model. Following numbers can illustrate (see Fig.~\ref{cc-rat}) 
the $E_c$ at INO depth for some of charm hadroproduction models: $E_c^{\rm GGV05}\simeq 250$ TeV,  $E_c^{\rm QGSM}\simeq 300$ TeV,  $E_c^{\rm GGV01}\simeq 600$ TeV. 
 
From the Table~\ref{t3} we can see that prompt muon flux contribution due to GGV01 model, for example, may differs from that for the GGV051 model (or QGSM)  by factor 2 at $E_\mu > 200$ TeV. In other words, expected number of muon events inside the INO detector may increase by      $50$~\% at the energy above $200$ TeV if GGV05 or QGSM predictions are  reasonable.

%%%%%%%%%%%%%%%%%%%%%%%%%%%%%%%%%%%%%%%%%%%%%%%%%%%%%%%%%%%%%%%%%%%%

\section {Acknowledgements}
The work of S.P. was supported by the Ministerio de Educación y Ciencia
under Proyecto Nacional FPA2006-01105, and also by the Comunidad de Madrid
under Proyecto HEPHACOS, Ayuda de I+D S-0505/ESP-0346. 
The author S.P would like to thank Indumati for 
providing valuable information regarding INO experiment. We thank Pankaj
jain for careful reading the manuscript.  
S.~Sinegovsky acknowledges the support by 
Federal Programme  "Leading Scientific Schools of 
Russian Federation",  grant NSh-5362.2006.2.

%%%%%%%%%%%%%%%%%%%%%%%%%%%%%%%%%%%%%%%%%%%%%%%%%%%%%%%%%%%%%%%%%%%
\pagebreak

\end{document}